\documentclass[sigconf]{acmart}
\AtBeginDocument{%
  }

\setcopyright{acmlicensed}
\copyrightyear{2018}
\acmYear{2018}
\acmDOI{XX.XXX}
\acmConference[Conference'XX]{xx}{June 03--05,
  2018}{Woodstock, NY}
\acmISBN{978-1-4503-XXXX-X/2018/06}

\usepackage{amsmath}
\usepackage{algorithm}
\usepackage{graphicx}
\usepackage{booktabs}
\usepackage{xcolor}
\usepackage{colortbl}
\usepackage{subcaption}
\usepackage{fancybox}
\usepackage[most]{tcolorbox}
\usepackage{algpseudocode}
\usepackage{amsmath}
\usepackage{amsthm}
\usepackage{multirow}
\usepackage{mathrsfs}
\usepackage{pgfplots}
\usepackage{soul,color}
\usepackage{wrapfig}
\usepackage[table]{xcolor}
\definecolor{lightblue}{RGB}{230,240,255}

\begin{document}

\title{A Generative Contextual Comprehension Paradigm for Takeout Ranking Model}


\author{Ziheng Ni\textsuperscript{*},Congcong Liu\textsuperscript{*\dag}, Cai Shang, Yiming Sun, Junjie Li, Zhiwei Fang, Guangpeng Chen, Li Jian, Zehua Zhang, Changping Peng, Zhangang Lin, Ching Law, Jingping Shao }

\affiliation{%
  \institution{JD.com}
  \country{Beijing, China}
  }
  

\renewcommand{\shortauthors}{Ni and Liu et al.}

\begin{abstract}

The ranking stage serves as the central optimization and allocation hub in advertising systems, governing economic value distribution through eCPM and orchestrating the user-centric blending of organic and advertising content. 
Prevailing ranking models often rely on fragmented modules and hand-crafted features, limiting their ability to interpret complex user intent.  
This challenge is further amplified in location-based services such as food delivery, where user decisions are shaped by dynamic spatial, temporal, and individual contexts.
To address these limitations, we propose a novel generative framework that reframes ranking as a context comprehension task, modeling heterogeneous signals in a unified architecture.
Our architecture consists of two core components: the Generative Contextual Encoder (GCE) and the Generative Contextual Fusion (GCF). The GCE comprises three specialized modules: a Personalized Context Enhancer (PCE) for user-specific modeling, a Collective Context Enhancer (CCE) for group-level patterns, and a Dynamic Context Enhancer (DCE) for real-time situational adaptation. The GCF module then seamlessly integrates these contextual representations through low-rank adaptation.
Extensive experiments confirm that our method achieves significant gains in critical business metrics, including click-through rate and platform revenue. 
We have successfully deployed our method on a large-scale food delivery advertising platform, demonstrating its substantial practical impact. 
This work pioneers a new perspective on generative recommendation and highlights its practical potential in industrial advertising systems.
\end{abstract}



\maketitle

\begingroup
\renewcommand\thefootnote{}
\footnotetext{\textsuperscript{*}Equal contribution}
\footnotetext{\textsuperscript{\dag}Corresponding author}
\endgroup

\section{Introduction}
\label{sec:introduction}

In the context of takeout advertising, ranking plays an essential role in determining ad exposure and allocation. The ranking scores generated during this stage, such as predicted Click-Through Rate (pCTR) and predicted Conversion Rate (pCVR), are integral components in the expected Cost Per Mille (eCPM) calculation. Within the eCPM framework, these predicted values are combined with the advertiser's bid price and other strategic factors reflecting both advertiser requirements and user experience considerations. Together, these elements jointly determine the final ranking and allocation of advertisements.
In addition, modern advertising platforms frequently interleave ads with organic content to enhance user experience. In food delivery scenarios, for instance, advertisements are seamlessly blended with organic search results or recommendation lists, necessitating a mixed ranking process. This demands a unified metric to evaluate the relative utility of each item, whether advertised or organic. Traditional fixed-position strategies prove inadequate in such settings, as they fail to adapt to the dynamic interplay between ad relevance, user intent, and business objectives. Consequently, industrial mixed-ranking systems require sophisticated modeling techniques capable of jointly optimizing for both advertiser value and user satisfaction.

As a typical Location-Based Service (LBS) scenario, takeout advertising must account for numerous unique spatiotemporal factors that significantly influence user experience and operational efficiency. These include strict spatiotemporal constraints (e.g., delivery distance and preparation time), user preferences for specific Points of Interest (POI) like offices or residential compounds, and broader Area of Interest (AOI) characteristics such as commercial districts or university towns. For instance, a user in a business district during lunchtime may prioritize restaurants with fast preparation and short delivery times, while the same user in a residential area during dinner might value variety over speed. Currently, generative approaches for LBS scenarios remain relatively under-explored. While OneLoc~\cite{oneloc} proposes an end-to-end item generation framework focusing on geographical context constraints for group-buying scenarios in local services, its primary emphasis on spatial proximity is insufficient for the more complex demands of takeout advertising. Beyond geographical constraints, the takeout scenario requires integrated consideration of multifaceted contextual signals particularly temporal factors (e.g., time-of-day, peak hours, real-time delivery capacity), which are crucial for accurate ranking and allocation. However, a systematic solution that comprehensively integrates these diverse contextual dimensions into a generative framework for takeout advertising remains an open challenge.

To address this gap, we propose \textbf{GCRank} (\textbf{G}enerative \textbf{C}ontextual \textbf{Rank}), a novel generative paradigm specifically designed for contextual comprehension in takeout advertising ranking. Our framework systematically decomposes the complex contextual environment into four distinct yet complementary dimensions: static context ( including restaurant attributes and fixed user profiles), dynamic context (encompassing real-time factors such as delivery capacity, weather conditions, and time-sensitive demands), personalized context (leveraging individual user preferences and behavioral history), and collective context (capturing group-level patterns and regional trends through AOI-aware modeling).
At the core of GCRank lies the \textbf{Generative Takeout Context Encoder (GCE)}, a unified architecture that employs advanced generative techniques to holistically model both the individual characteristics and, more importantly, the complex interrelationships among these contextual dimensions. The GCE comprises three specialized modules: a \textbf{Personalized Context Enhancer (PCE)} for user-specific preference modeling, a \textbf{Collective Context Enhancer (CCE)} for mining group-level semantic patterns, and a \textbf{Dynamic Context Enhancer (DCE)} for real-time situational reasoning. Through these modules, GCRank learns to generate unified context representations that effectively capture intricate dependencies between spatial constraints, temporal dynamics, personal preferences, and collective behaviors. 
For instance, our model can precisely characterize how a user's food preference (personalized context) interacts with current delivery capacity (dynamic context) during rainy weather (dynamic context) in a specific business district (static and collective context). This holistic contextual understanding, further refined through our \textbf{Generative Context Fusion (GCF)} module, enables GCRank to produce more accurate and context-aware ranking scores, specifically tailored to the unique challenges of takeout advertising scenarios.

The main contributions of this work are summarized as follows:
\begin{itemize}
    \item \textbf{Novel Generative Paradigm:} We propose the first generative context understanding paradigm specifically designed for takeout advertising ranking, systematically addressing the unique challenges of LBS-aware advertising through a unified contextual comprehension framework.
    
    \item \textbf{Specialized Context Modeling Architecture:} We design a comprehensive architecture with three core modules for specialized context processing: a Collective Context Enhancer (CCE) for mining group-level patterns and regional trends, a Dynamic Context Enhancer (DCE) for modeling real-time contextual factors, and a Personalized Context Enhancer (PCE) for capturing individual user preferences with spatiotemporal adaptation.
    
    \item \textbf{Advanced Context Fusion Methodology:} We introduce a novel Generative Context Fusion (GCF) method that effectively addresses the alignment and integration challenges of heterogeneous contextual information through low-rank adaptation and cross-attention mechanisms.
    
    \item \textbf{Comprehensive Experimental Validation:} We conduct extensive experiments on industrial datasets, demonstrating significant improvements over traditional methods and recent generative solutions. Ablation studies further validate the contribution of each proposed component, while comparative experiments on public datasets confirm the generalizability of our approach.
    
    \item \textbf{Industrial Impact and Resource Contribution:} Our method has been successfully deployed on a large-scale takeout advertising platform, achieving remarkable improvements. Recognizing the limitations of existing public datasets for generative research, we will release an anonymized industrial dataset to facilitate future work in this domain.
\end{itemize}

\section{Related Work}
\label{sec:related_work}
\subsection{Classical Ranking Models}
Traditional ranking models in advertising, particularly the Deep Learning Recommendation Model (DLRM) architecture, typically process multiple input features including contextual information (e.g., time, location), user profiles (e.g., gender, age), user behavior sequences, and various cross-features of target items. Two particularly important modules in ranking models are behavior sequence processing and feature interaction learning.
Behavior sequence modules~\cite{zhou2018deep, zhou2019deep,pi2020search,ETA,twin,twinv2,SDIM} commonly employ target attention mechanisms to capture similarities between users' historical behaviors and candidate items. These approaches effectively model users' dynamic preferences by attending to relevant historical interactions. Meanwhile, feature interaction modules~\cite{cheng2016wide,guo2017deepfm,xdeepfm,dcn,dcnv2} aim to capture complex interactions among different features, including user and item characteristics, to generate final predictions. These methods typically utilize various interaction operations such as inner products, cross networks, or self-attention mechanisms to model feature correlations.

\subsection{Generative Recommendation}
Recently, generative recommendation~\cite{hstu,mtgr,deng2025onerec,onerec_tech,onerecv2,onesearch,onesug,oneloc,unisearch,lum,hllm,onepiece,egav2} has attracted significant attention from both academia and industry. Methods such as EGA-V2~\cite{egav2}, OneRec~\cite{deng2025onerec, onerec_tech, onerecv2}, OneSug~\cite{onesug}, OneSearch~\cite{onesearch}, UniSearch~\cite{unisearch}, and OneLoc~\cite{oneloc} focus on end-to-end reconstruction to improve both effectiveness and efficiency, while incorporating reward designs to introduce additional signals such as ranking scores and geographical rewards for user preference learning.
Other approaches like MTGR~\cite{mtgr} and HSTU~\cite{hstu} transform traditional DLRM by serializing all features in temporal order. However, these implementations remain essentially discriminative in nature for ranking tasks. More importantly, existing generative methods lack systematic modeling of contextual information in ranking scenarios and fail to consider the distinct semantic meanings of different contextual dimensions. Furthermore, the generative paradigm for contextual understanding in takeout advertising remains largely unexplored, particularly in handling the complex interplay between spatial, temporal, personal, and collective contextual factors that are unique to LBS-based food delivery scenarios.

\section{Preliminary}
\label{sec:preliminary}
In this section, we formalize the ads ranking problem in food delivery platforms. We consider the complete environmental information available at the time of a user request as the contextual basis for our prediction model. The contextual information is decomposed into four distinct but complementary dimensions: static context, dynamic context, personalized context, and collective context~\cite{onepiece}.

\subsection{Contextual Taxonomy}
The contextual information $C$ for a user request is represented as a quadruple $C = (C_s, C_d, C_p, C_c)$, where each component captures a specific aspect of the environment.
The \textbf{static context} $C_s$ encompasses slowly-changing information that remains stable over extended periods. Formally, we represent $C_s$ as a feature vector $\mathbf{c}_s \in \mathbb{R}^{d_s}$ that encodes user attributes (e.g., gender, age), store attributes (e.g., category), and environmental constants (e.g., season).
The \textbf{dynamic context} $C_d$ captures real-time information that varies with each request, directly influencing immediate user intent. This component is represented as $\mathbf{c}_d \in \mathbb{R}^{d_d}$, incorporating temporal signals (e.g., hour), spatial coordinates, and environmental conditions (e.g., weather, temperature).
The \textbf{personalized context} $C_p$ encodes user-specific behavioral patterns through historical interaction sequences. For user $u$, we represent this as $S_u = \{(a_i, \tau_i)\}_{i=1}^n$ where $a_i$ denotes an ad interaction, and $\tau_i$ is the additional features of the behavior.
The \textbf{collective context} $C_c$ captures population-level behavioral patterns by explicitly modeling AOI-aware information and group dynamics. This component is formalized as $\mathbf{c}_c \in \mathbb{R}^{d_c}$, which is derived from statistical aggregates of user interactions within specific geographical and contextual boundaries. It particularly emphasizes the mining of AOI-aware signals, including regional preference patterns, area-specific popularity trends (e.g., trending merchants and categories within particular AOIs).

\subsection{Problem Formulation}
Given a user $u$, a candidate set of ads items $\mathcal{I} = \{i_1, i_2, \dots, i_N\}$, and the complete context $C = (C_s, C_d, C_p, C_c)$, our goal is to learn a scoring function $f: \mathcal{I} \times \mathcal{C} \rightarrow \mathbb{R}$ that estimates the probability of user engagement with ad $i_j$ under context $C$. Formally, we define:
\begin{equation}
    f(i_j, C) = \mathbb{P}(y = 1 \mid i_j, C; \Theta)
\end{equation}
where $y \in \{0, 1\}$ is the binary engagement indicator, and $\Theta$ represents the model parameters.

\section{Methodology}
\label{sec:methodology}
\begin{figure*}[t]
    \centering
    \includegraphics[width=\textwidth]{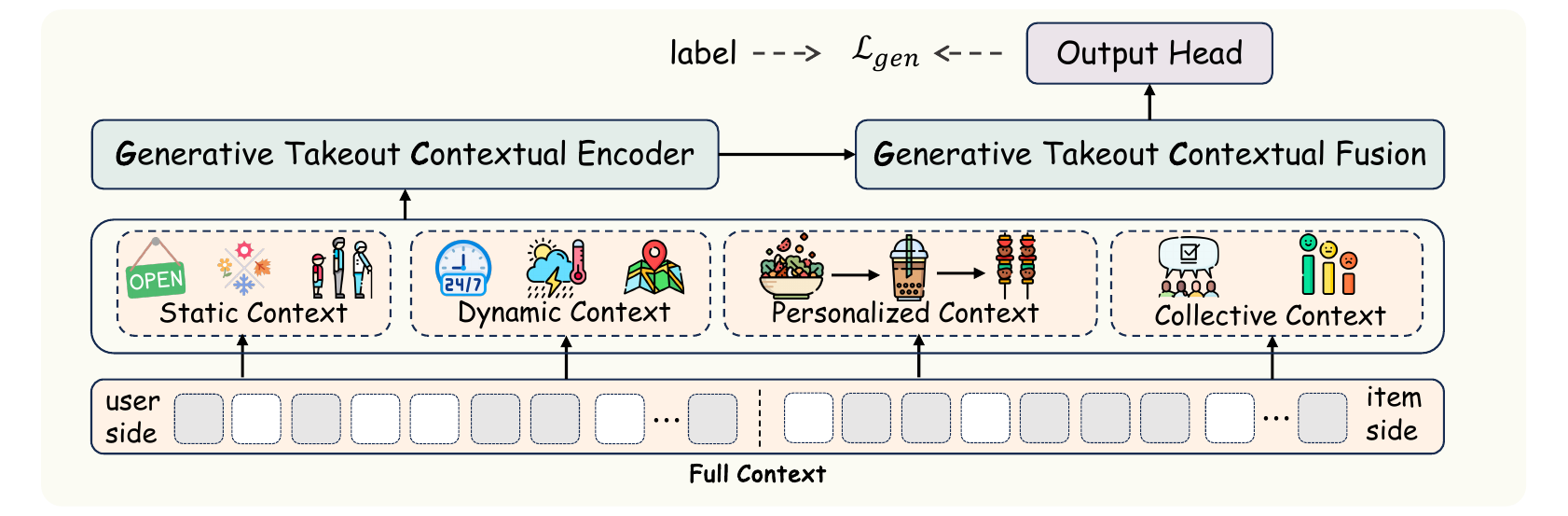}
    \caption{The overall framework of GCRank, consisting of two core modules: GCE (Generative Takeout Contextual Encoder) and GCF (Generative Contextual Fusion).}
    \label{fig:main}
\end{figure*}

In this section, we present the detailed architecture of our proposed \textbf{GCRank} framework. As illustrated in Figure~\ref{fig:main}, GCRank is designed to comprehensively model the complex contextual signals in takeout advertising through two core components: the \textbf{Generative Takeout Contextual Encoder (GCE)} and the \textbf{Generative Contextual Takeout Fusion (GCF)} Module. The GCE component specializes in capturing and encoding four distinct types of contextual information, static, dynamic, personalized, and collective contexts that are crucial for accurate ranking in LBS-based food delivery scenarios. 
Specifically, GCE is composed of three specialized submodules: the Personalized Context Enhancer (PCE) for modeling user-specific preferences, the Collective Context Enhancer (CCE) for extracting group-level semantic patterns, and the Dynamic Context Enhancer (DCE) for real-time situational reasoning. 
Subsequently, the GCF Module effectively integrates these heterogeneous contextual representations, learning the intricate relationships and dependencies between different context dimensions to generate unified and discriminative features for final ranking prediction. In the following subsections, we describe each module in detail.

\begin{figure*}[t]
    \centering
    \includegraphics[width=\textwidth]{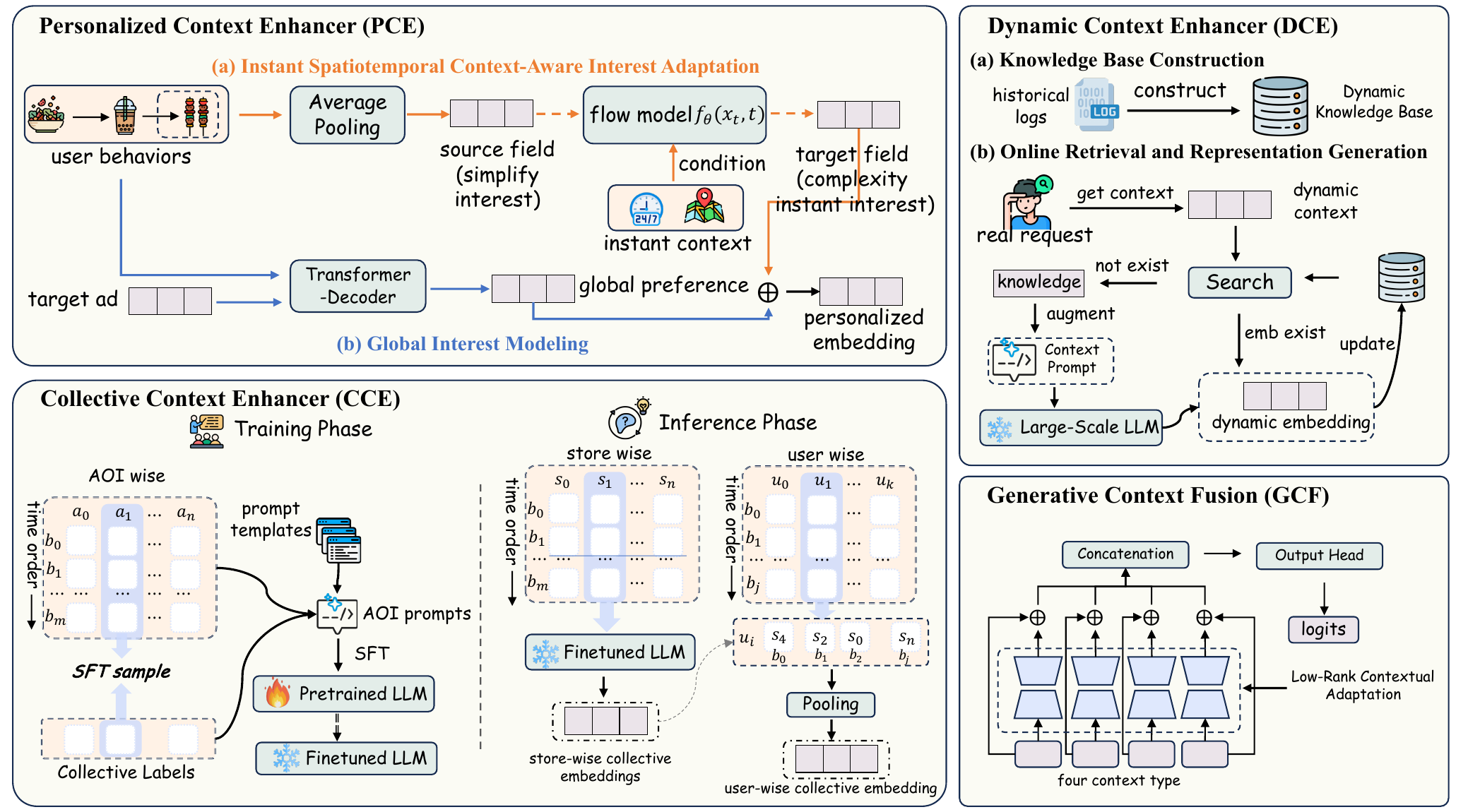}
    \caption{The detailed architecture of GCE and GCF.GCE is composed of three sub-modules: Personalized Context Enhancer (PCE), Dynamic Context Enhancer (DCE), and Collective Context Enhancer (CCE).}
    \label{fig:detail}
\end{figure*}

\begin{algorithm}[t]
\caption{Online Inference of DCE Module}
\label{alg:dce}
\begin{algorithmic}[1]
\Require Dynamic context $c_q$, Knowledge base $\mathcal{K}$, Embedding model $g_{\phi}$, LLM $G_{\theta}$, similarity threshold $\tau$
\Ensure Dynamic context-aware embedding $\mathbf{e}_d$
\State $\mathbf{k}_q \gets g_{\phi}(c_q)$
\State $(\mathbf{k}^*, c^*, s^*, \mathbf{e}_d^*) \gets \arg\max_{(\mathbf{k}_i, s_i) \in \mathcal{K}} \text{sim}(\mathbf{k}_q, \mathbf{k}_i)$
\If {$\text{sim}(\mathbf{k}_q, \mathbf{k}^*) > \tau$}
    \State \Return $\mathbf{e}_d^*$ \Comment{Cache hit: return precomputed embedding}
\Else
    \State $\mathcal{P}_d \gets \text{FormatPrompt}(c_q, s^*)$ \Comment{Cache miss: call LLM}
    \State $\mathbf{H} \gets G_{\theta}(\mathcal{P}_d)$ \Comment{Get hidden states from LLM}
    \State $\mathbf{e}_d^* \gets \text{AveragePooling}(\mathbf{H})$
    \State $\mathcal{K} \gets \mathcal{K} \cup \{(\mathbf{k}_q, c_q, s^*,\mathbf{e}_d^*)\}$ \Comment{Asynchronous update}
    \State \Return $\mathbf{e}_d^*$
\EndIf
\end{algorithmic}
\end{algorithm}

\subsection{GCE: Dynamic Context Enhancer}
\label{subsec:dynamic_context}
The \textbf{Dynamic Context Enhancer (DCE)} module is designed to capture and model highly volatile contextual signals in real-time advertising requests. The module processes dynamic context features including temporal, spatial, weather, and immediate user signals through a structured knowledge retrieval and reasoning framework.

\subsubsection{Knowledge Base Construction}
\label{subsubsec:knowledge_construct}
We construct a searchable knowledge base $\mathcal{K}$ from aggregated historical logs, encoding observed user preferences and system behaviors under diverse contextual conditions. Each dynamic context $c_i \in C_d$ is represented as a feature tuple (e.g., $c_i = (\text{hour}=\text{``19:00''}, \text{location}=\text{``CBD''}, \text{weather}=\text{``rainy''}, \text{holiday}=\text{``Winter Solstice''})$). For each unique context, we compile a structured knowledge snippet $s_i$ containing popular categories and merchant preferences, delivery time and fee estimates, price and promotion sensitivity metrics, as well as seasonal and holiday-specific patterns.
The knowledge base is organized as a quadruple:
\begin{equation}
    \mathcal{K} = \{ (\mathbf{k}_i, c_i, s_i, \mathbf{e}_{\text{d}}^i) \mid i=1, ..., N \}
\end{equation}
where $\mathbf{k}_i = g_{\phi}(c_i)$ denotes the dense vector representation of context $c_i$ obtained via embedding model $g_{\phi}$ such as BGE~\cite{bge}, $s_i$ represents the structured knowledge snippet for context $c_i$, $\mathbf{e}_{d}^i$ is the precomputed context-aware embedding generated by processing $s_i$ through our LLM-based reasoning pipeline.
During knowledge base construction, for each context $c_i$, we generate the corresponding $\mathbf{e}_d^i$ by formatting $s_i$ into a structured prompt and processing it through the Large Language Model $G_{\theta}$, followed by the same average pooling operation described in Section~\ref{subsubsec:dynamic_online}.

\subsubsection{Online Retrieval and Representation Generation}
\label{subsubsec:dynamic_online}
For an incoming request with context $c_q$, we compute its embedding $\mathbf{k}_q = g_{\phi}(c_q)$ and retrieve the most relevant knowledge entry through maximum cosine similarity search:
\begin{equation}
    (\mathbf{k}^*, c^*, s^*, \mathbf{e}_d^*) = \arg\max_{(\mathbf{k}_i, c_i, s_i, \mathbf{e}_d^i) \in \mathcal{K}} \frac{\mathbf{k}_q \cdot \mathbf{k}_i}{\|\mathbf{k}_q\| \|\mathbf{k}_i\|}
\end{equation}
The retrieved context-aware embedding $\mathbf{e}_d^*$ is directly utilized as the dynamic context representation. For novel contexts that cannot be matched with sufficient similarity, we trigger the full LLM inference pipeline: the retrieved knowledge $s^*$ is formatted into a structured prompt $\mathcal{P}_{d}$ and processed by $G_{\theta}$, generating the context-aware embedding via average pooling of the final layer hidden states $\mathbf{H} \in \mathbb{R}^{N \times d}$:
\begin{equation}
    \mathbf{e}_d = \frac{1}{N} \sum_{t=1}^{N} \mathbf{H}_t
\end{equation}

\subsubsection{Optimized Inference Protocol}
\label{subsubsec:dynamic_infer}
We implement a cache-based inference optimization where the current context embedding $\mathbf{k}_q$ is first matched against the knowledge base. If a sufficiently similar context exists ($\text{sim}(\mathbf{k}_q, \mathbf{k}_i) > \tau$), the precomputed embedding $\mathbf{e}_d^i$ is directly utilized without LLM inference. Otherwise, the full LLM inference pipeline is triggered and the knowledge base is asynchronously updated with the new quadruple $(\mathbf{k}_q, c_q, s_q, \mathbf{e}_d)$.
The knowledge base undergoes weekly comprehensive updates using recent impression logs, while the hot-update mechanism handles novel contexts in real-time, as formalized in Algorithm~\ref{alg:dce}.

\subsection{GCE: Collective Context Enhancer}
\label{subsec:collective_context}
The \textbf{Collective Context Enhancer (CCE)} module generates semantic representations through AOI-aware collective learning. We employ a supervised fine-tuning (SFT) approach at the AOI granularity, using collective attributes including taste preferences, consumption levels, and user demographics as training labels. The trained model subsequently enables inference at both store and user granularities.

\subsubsection{Training Phase: AOI-Granularity Supervision}
Let $\mathcal{A}$ denote the set of AOIs. For each AOI $a \in \mathcal{A}$, we aggregate chronological click sequences from all constituent stores over 60 days:
$$
    B_a = \{b_1, b_2, \dots, b_N\}
$$
where each click event $b_i$ is enriched with user attributes (e.g., demographic features, historical preferences) and spatiotemporal context. 
We concatenate all behavioral information from $B_a$ according to predefined formatting rules to form a comprehensive training sample. This consolidated representation is formatted using prompt template $\mathcal{P}_c^{\text{train}}$.
The collective label $\mathbf{y}_a$ for AOI $a$ is constructed from aggregated attributes including taste preference distributions (e.g., spicy, sweet, savory), consumption level indicators (e.g., average order value, price sensitivity), user demographic profiles (e.g., age groups, occupational distributions), and temporal consumption patterns (e.g., peak hours, seasonal variations).

\subsubsection{Inference Phase: Multi-Granularity Generation}
\textbf{Store-level Semantic Representation}. For store $s$ in AOI $a$, we consolidate its recent click sequence $B_s^{\text{inf}}$ into a single comprehensive input using the same concatenation and formatting rules employed during training. The store embedding is generated by processing this unified representation through the fine-tuned LLM and applying average pooling over the final layer hidden states:
\begin{equation}
    \mathbf{H}_s^{\text{store}} = \text{LLM}(\text{Concat-Format}(B_s^{\text{inf}}, \mathcal{P}_c^{\text{inf}}))
\end{equation}
\begin{equation}
    \mathbf{e}_s^{\text{store}} = \frac{1}{N_c} \sum_{n=1}^{N_c} \mathbf{H}_{s,n}^{\text{store}}
\end{equation}

\textbf{AOI-level Semantic Representation}. The AOI embedding $\mathbf{e}_a^{\text{AOI}}$ is generated by applying the same concatenation and formatting procedure to the complete behavioral set $B_a$ from the target AOI, then processing through the fine-tuned LLM followed by average pooling.

\textbf{User-level Collective Profile}. A user $u$'s collective representation integrates store-level embeddings from their interaction history, weighted by AOI context:
\begin{equation}
    \mathbf{e}_u = \frac{1}{|S_u|} \sum_{s \in S_u} (\beta \cdot \mathbf{e}_s^{\text{store}} + (1-\beta) \cdot \mathbf{e}_{A(s)}^{\text{AOI}})
\end{equation}

where $S_u$ represents stores clicked by user $u$, $A(s)$ denotes the AOI of store $s$, and $\beta$ balances store-specific and AOI-aware signals.

These multi-granularity semantic representations $\{\mathbf{e}_s^{\text{store}}, \mathbf{e}_a^{\text{AOI}}, \mathbf{e}_u\}$ capture collective consumption patterns at different abstraction levels for downstream ranking.

\begin{algorithm}[t]
\caption{Collective Semantic Generation with AOI Supervision}
\label{alg:collective}
    \begin{algorithmic}[1]
        \Procedure{Training}{}
        \For{each AOI $a \in \mathcal{A}$}
        \State Aggregate chronological clicks $B_a$ from all stores in $a$
        \State Extract collective label $\mathbf{y}_a$
        \State Format consolidated training sample $\mathbf{x}_a \gets \text{Concat-Format}(B_a, \mathcal{P}_{\text{train}})$
        \State Add $(\mathbf{x}_a, \mathbf{y}_a)$ to $\mathcal{D}_{\text{train}}$
        \EndFor
        \State Fine-tune LLM on $\mathcal{D}_{\text{train}}$ with collective supervision
        \EndProcedure
        \Procedure{Inference}{}
        \For{each store $s \in S$}
        \State Consolidate recent clicks $B_s^{\text{inf}}$ into unified input
        \State Compute $\mathbf{H}_s^{\text{store}} \gets \text{LLM}(\text{Concat-Format}(B_s^{\text{inf}}, \mathcal{P}_{\text{inf}}))$
        \State Compute $\mathbf{e}_s^{\text{store}} \gets \text{AveragePool}(\mathbf{H}_s^{\text{store}})$
        \EndFor
        \For{each AOI $a \in \mathcal{A}$}
        \State Compute $\mathbf{H}_a^{\text{AOI}} \gets \text{LLM}(\text{Concat-Format}(B_a, \mathcal{P}_{\text{inf}}))$
        \State Compute $\mathbf{e}_a^{\text{AOI}} \gets \text{AveragePool}(\mathbf{H}_a^{\text{AOI}})$
        \EndFor
        \For{each user $u \in U$}
        \State $\mathbf{e}_u \gets \frac{1}{|S_u|} \sum_{s \in S_u} (\beta \cdot \mathbf{e}_s^{\text{store}} + (1-\beta) \cdot \mathbf{e}_{A(s)}^{\text{AOI}})$
        \EndFor
        \EndProcedure
    \end{algorithmic}
\end{algorithm}

\subsection{GCE: Personalized Context Enhancer}
\label{subsec:personalized_context}
The \textbf{Personalized Context Enhancer (PCE)} module addresses the highly dynamic nature of user preferences in food delivery scenarios, where interests exhibit significant variations across different times of day and geographical locations. The PCE comprises two complementary components: global personalized interest modeling and instant spatiotemporal context-aware interest adaptation.

\subsubsection{Global Personalized Interest Modeling}
We first extract the user's long-term preference patterns from their historical behavior sequence. Let $H_p^u = \{b_1, b_2, \dots, b_T\}$ represent the chronological sequence of user $u$'s historical behaviors, where each behavior $b_t$ includes interacted items and corresponding contextual features. We employ a transformer-based sequence modeling module with target attention to encode this sequence:
\begin{equation}
    \mathbf{e}_{p,u}^{\text{global}} = \text{Transformer-Decoder}(H_p^u, target\_ad)
\end{equation}
The target attention mechanism enables the model to adaptively focus on historical behaviors that are most relevant to the current candidate item, enhancing the relevance of the generated representations. This global user embedding $\mathbf{e}_{p,u}^{\text{global}}$ captures stable preference patterns across different contexts and serves as the foundation for personalized ranking.

\subsubsection{Instant Spatiotemporal Context-Aware Interest Adaptation}
To address the sparsity issue under fine-grained spatiotemporal context partitioning, we reformulate instant interest modeling as a distribution transformation process rather than knowledge co-occurrence mining. Specifically, we leverage the flow matching paradigm to reshape the interest distribution according to immediate contextual signals.
Let $\mathbf{e}_{p,u}^{\text{avg}}$ denote the user's average interest representation derived from their historical behaviors excluding the last interaction, serving as the source distribution. For a given instant context $c_{\text{ins}}$ (including precise time and location information), we construct a guidance signal $\mathbf{g}_{\text{ins}}$ that encodes the fine-grained spatiotemporal characteristics.
To optimize the quality of flow matching generation, we employ the embedding of the user's last historical interaction $\mathbf{e}_{\text{last}}$ as the ground truth target distribution, while using the corresponding instant context of that last interaction as the guidance signal during training. This ensures that the generated interest representations accurately reflect the user's preferences under specific contextual conditions.
The transformation from source to target distribution is governed by a velocity field $v_\theta$ parameterized by a neural network:
\begin{equation}
    \frac{d\mathbf{e}_{p,u}(t)}{dt} = v_\theta(\mathbf{e}_{p,u}(t), \mathbf{g}_{\text{ins}}, t)
\end{equation}
where $\mathbf{e}_{p,u}(0) = \mathbf{e}_{p,u}^{\text{avg}}$ and $\mathbf{e}_{p,u}(1)$ represents the final context-aware user interest.
To enhance inference efficiency, we adopt the MeanFlow~\cite{geng2025mean} optimization strategy, which approximates the continuous transformation through direct integration of the velocity field.

\subsubsection{Personalized Interest Fusion}
The final personalized user representation combines both global and instant context-aware interests through direct concatenation:
\begin{equation}
    \mathbf{e}_{p}^{u} = [\mathbf{e}_{p,u}^{\text{global}}; \mathbf{e}_{p,u}^{\text{instant}}]
\end{equation}

\subsection{Generative Contextual Fusion}
\label{subsec:context_fusion}
We transform each contextual representation into a shared semantic space through parameter-efficient low-rank adapters. For each context type $i \in \{s,d,p,c\}$, we apply a LoRA-style transformation:
\begin{equation}
    \mathbf{e}_i' = \mathbf{e}_i + \mathbf{B}_i\mathbf{A}_i\mathbf{e}_i
\end{equation}
where $\mathbf{A}_i \in \mathbb{R}^{r \times d_i}$ and $\mathbf{B}_i \in \mathbb{R}^{d \times r}$ are low-rank matrices with rank $r \ll d_i$. We employ adaptive rank configuration based on context complexity: dynamic and personalized contexts utilize large $r$ for fine-grained adaptation, while static and collective contexts employ small $r$ due to their more stable patterns. This design achieves parameter efficiency scaling as $\mathcal{O}(r \times (d_i + d))$ per context type while maintaining representational capacity through residual connections.

The aligned representations are then concatenated and processed through a multi-layer perceptron to generate the final fused representation:
\begin{equation}
    \mathbf{e}_{\text{fusion}} = \text{MLP}([\mathbf{e}_s'; \mathbf{e}_d'; \mathbf{e}_p'; \mathbf{e}_c'])
\end{equation}
This streamlined fusion approach effectively integrates the complementary information from different contextual dimensions while maintaining computational efficiency for real-time ranking.

\section{Experiment}
\label{sec:experiment}
\subsection{Experiment Settings}
\label{subsec:exp_settings}
\subsubsection{Datasets}
\label{subsubsec:datasets}
While publicly available datasets for recommendation and advertising tasks especially in takeout scenario predominantly utilize numerical features, they often lack the rich textual semantic information that is crucial for our generative contextual comprehension paradigm. Plain-text data provides more nuanced and expressive signals compared to numerical features, serving as the fundamental input for our method. To address the scarcity of semantic information in existing public datasets, we construct a comprehensive training dataset based on logs from a real-world industrial takeout advertising platform~\footnote{To safeguard proprietary business information, we applied carefully designed transformations to the reported results. These transformations preserve the essential statistical properties while ensuring that no sensitive business-related data can be reverse-engineered from the published outcomes.}. 
It contains about 2 billions of samples, covering a large user base and several millions of ads.

\subsubsection{Baselines}
\label{subsubsec:baselines}
We compare GCRank against several state-of-the-art methods spanning different paradigms: DNN as the fundamental deep learning baseline; DIN~\cite{zhou2018deep} with adaptive interest modeling; SIM~\cite{sim} for long sequence modeling; HSTU~\cite{hstu} representing a leading proposal for generative ranking reformulation; and OneLoc~\cite{oneloc} as the recent generative approach for LBS scenarios. Additionally, we include our production-level DLRM model as a strong industrial baseline, which has undergone multiple rounds of optimization and represents the current state-of-the-art in our real-world advertising system. For OneLoc, while it is designed as an end-to-end solution, we adapt its geographical context utilization methodology for fair comparison in our offline experiments, focusing on its core innovations for LBS scenarios. This comprehensive set of baselines ensures rigorous evaluation across different architectural paradigms and practical industrial requirements.

\subsubsection{Evaluation Metrics}
\label{subsubsec:metrics}
In our offline evaluation, we focus on two key tasks: CTR (Click-Through Rate) and CVR (Conversion Rate) prediction. We employ AUC (Area Under the ROC Curve) as our primary evaluation metric, where an improvement of 0.0001 is considered statistically significant in industrial advertising scenarios. Additionally, we introduce the RelaImpr metric to quantify relative improvements over baseline models following ~\cite{zhou2018deep,relaimpr1,diffumin}. RelaImpr is defined as:
\begin{equation}
    \text{RelaImpr}=(\frac{\text{AUC(model)}-0.5}{\text{AUC(base  model)}-0.5}-1) \times 100\%
\end{equation}

\subsubsection{Implement Details}
\label{subsubsec:implement}
Our model is trained on NVIDIA A100 GPUs using the Adam optimizer with a learning rate of 0.00002 and batch size of 512. 
We utilize user behavior sequences from the past 60 days, truncated to a maximum length of 300 interactions. 
For the CCE Component, we initialize with Qwen1.5-0.5B as the pretrained LLM backbone, while the DCE utilizes the larger DeepSeek-R1 model for contextual reasoning. 
In the PCE module, we configure the CFG scale $\boldsymbol{\omega}=2.0$ and employ a lognormal sampler with $\boldsymbol{\mu}=-0.4$ and $\boldsymbol{\sigma}=1.2$ for the flow matching process.

\subsection{Overall Performance Comparison}
\label{subsec:exp_compare}
We evaluate the performance of GCRank against all baseline methods using the industrial dataset described in Section~\ref{subsubsec:datasets}. All results are the average of five repeated experiments. The comprehensive results are presented in Table~\ref{tab:main_results}. Following established practices in industrial recommendation systems, an AUC improvement of 0.001 is considered practically significant.
Our experimental results demonstrate that GCRank achieves state-of-the-art performance on both CTR and CVR prediction tasks, significantly outperforming all baseline methods. 

Among the comparative methods, HSTU shows competitive performance by incorporating contextual information as side features to enrich behavior representations, achieving better results than traditional sequential models. However, its improvement over the strongest sequential baselines remains limited, suggesting constraints in handling the complex contextual interactions characteristic of takeout advertising.

Interestingly, OneLoc, which focuses primarily on geographical context modeling, demonstrates limited effectiveness in our scenario. We hypothesize that this is because takeout advertising requires the integration of multiple contextual dimensions beyond just geographical information, including temporal dynamics, personalized preferences, and collective patterns. The single-dimensional enhancement provided by OneLoc proves insufficient to capture the rich contextual dependencies essential for optimal performance in this domain.
Notably, GCRank also significantly outperforms our production-level DLRM model, which has undergone extensive optimization and represents a strong industrial benchmark. This demonstrates the practical value of our generative contextual comprehension paradigm in real-world advertising systems.

These findings validate our key insight that a comprehensive generative approach that systematically integrates multiple contextual dimensions is crucial for achieving superior performance in complex LBS advertising scenarios like takeout platforms.
\begin{table}[h]
\caption{Overall performance comparison. Best results are in \textbf{bold}, second best results are \underline{underlined}. All results run over five times with std $\approx$ 1e-3.}
\label{tab:main_results}
\centering
\begin{tabular}{@{}lcccc@{}}
\toprule
\multirow{2}{*}{Method} & \multicolumn{2}{c}{CTR} & \multicolumn{2}{c}{CVR} \\
\cmidrule(lr){2-3} \cmidrule(lr){4-5}
 & AUC & RelaImpr & AUC & RelaImpr \\
\midrule
AvgPooling-DNN & 0.7403 & 0.00\% & 0.6503 & 0.00\% \\
DIN & 0.7462 & +2.45\% & 0.6562 & +3.93\% \\
SIM & 0.7481 & +3.25\% & 0.6588 & +5.66\% \\
HSTU & \underline{0.7545} & \underline{+5.91\%} & \underline{0.6659} & \underline{+10.38\%} \\
OneLoc & 0.7516 & +4.70\% & 0.6620 & +7.78\% \\
\midrule
Production DLRM & 0.7513 & +4.58\% & 0.6619 & +7.72\% \\
\midrule
\rowcolor{blue!10}
\textbf{GCRank (Ours)} & \textbf{0.7679} & \textbf{+11.49\%} & \textbf{0.7053} & \textbf{+36.59\%} \\
\bottomrule
\end{tabular}
\end{table}

\subsection{Ablation Study}
\label{subsec:exp_ablation}

To validate the effectiveness of each component in GCRank, we conduct comprehensive ablation studies on four core modules: \textbf{Dynamic Context Enhancer (DCE)}, \textbf{Collective Context Enhancer (CCE)}, \textbf{Personalized Context Enhancer (PCE)}, and \textbf{Generative Contextual Fusion (GCF)}. 

For the ablation experiments on DCE, CCE, and PCE, we remove the corresponding generative modules while incorporating the contextual information through standard feature integration approaches. Specifically, the contextual signals that would have been processed by these modules are instead introduced as conventional input features to maintain comparable information availability. For the GCF ablation, we remove the fusion module entirely and replace it with standard concatenation of context representations. The ablation results presented in Table~\ref{tab:ablation_result} demonstrate that removing any of these modules leads to significant performance degradation in both CTR and CVR tasks. This consistent performance decline across all ablated configurations confirms the essential role of our specialized generative modules in effectively modeling different contextual dimensions.

We further investigate the constituent sub-modules within CCE and PCE to understand their individual contributions. For CCE, we systematically ablate both the store-wise and user-wise embedding components. The detailed results reveal substantial performance drops in both prediction tasks, indicating that both embedding types play crucial and complementary roles in collective context understanding. Similarly, for PCE, we examine the Global Interest Modeling branch and the Instant Context-Aware Interest Adaptation branch separately. The observed performance decline underscores the complementary nature of these components: global preferences establish stable user profiling while dynamic adaptation captures immediate contextual influences, with both being indispensable for comprehensive personalization.

These comprehensive ablation studies systematically verify the contribution of each proposed module and sub-module, providing valuable insights into the architectural design rationale of GCRank and affirming the necessity of our holistic generative approach to contextual modeling.

\begin{table}[htbp]
\caption{Ablation studies of GCRank. All results run over five times with std $\approx$ 1e-3.}
\label{tab:ablation_result}
\centering
\begin{tabular}{@{}lcccc@{}}
\toprule
\multirow{2}{*}{Method} & \multicolumn{2}{c}{CTR} & \multicolumn{2}{c}{CVR} \\
\cmidrule(lr){2-3} \cmidrule(lr){4-5}
 & AUC & RelaImpr & AUC & RelaImpr \\
\midrule
    \textbf{GCRank} & \textbf{0.7679} & \textbf{0.00\%} & \textbf{0.7053} & \textbf{0.00\%} \\
    w/o DCE & 0.7626 & -1.98\% & 0.6898 & -7.55\% \\
    w/o CCE & 0.7621 & -2.16\% & 0.6920 & -6.48\% \\
    w/o PCE & 0.7652 & -1.01\% & 0.6971 & -3.99\% \\
    w/o GCF & 0.7643 & -1.34\% & 0.6967 & -4.19\% \\
\bottomrule
\end{tabular}
\end{table}

\subsection{Hyperparameter Experiments}
\label{subsec:exp_hyper}
\begin{figure}[t]
    \centering
    \includegraphics[width=\linewidth]{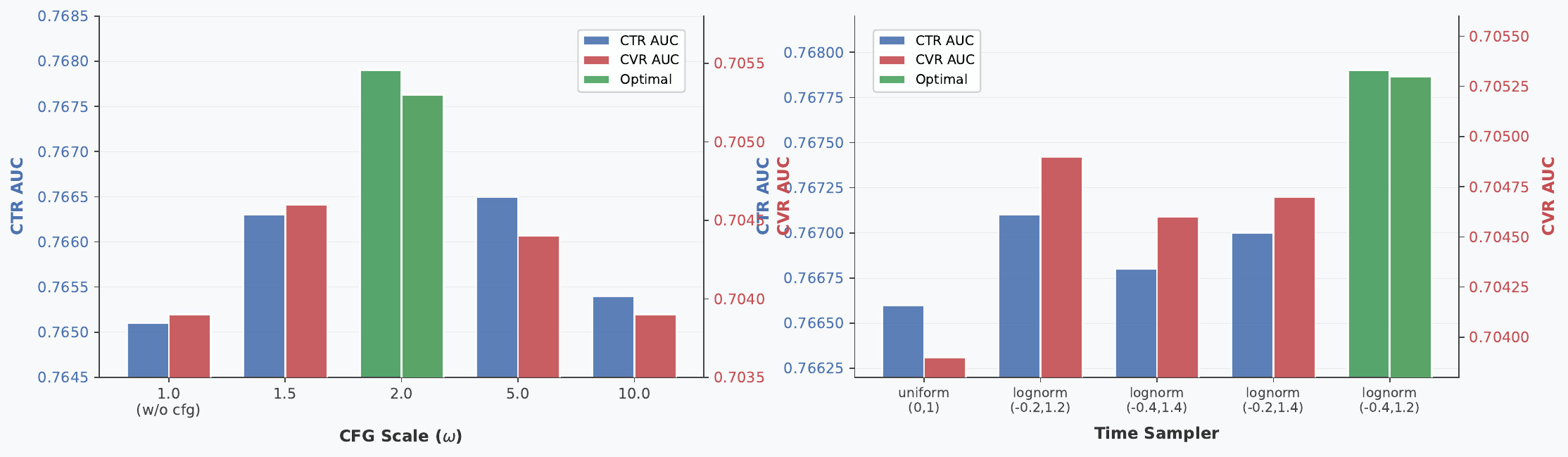}
    \caption{Hyperparameter studies of GCRank showing CTR AUC (left) and CVR AUC (right) performance across different configurations. Optimal configurations are highlighted in green.}
    \label{fig:hyperparameter_results}
\end{figure}

We conduct extensive hyperparameter experiments to thoroughly understand the sensitivity of our GCRank framework. Given that the PCE module introduces the most complex generative components specifically the conditional flow matching paradigm with multiple critical parameters, we focus our analysis primarily on its hyperparameters, including the CFG scale coefficient, and timestep sampling strategies. The comprehensive results are visualized in Figure~\ref{fig:hyperparameter_results}.
We investigate the Classifier-Free Guidance (CFG) scale coefficient $\omega$ in the PCE module. Our experiments reveal a relationship between $\omega$ and model performance, with optimal results achieved at $\omega = 2.0$. 

Meanwhile, building upon established findings~\cite{geng2025mean,esser2024scaling} that the distribution for sampling $t$ significantly influences generation quality, we systematically evaluate different sampling strategies for $(r, t)$ pairs in our PCE module. Our experimental setup employs independent sampling of $(r, t)$ followed by post-processing steps that enforce temporal ordering ($t > r$) through swapping operations and constrain the proportion of $r = t$ cases within specified bounds. Consistent with observations in Flow Matching~\cite{geng2025mean,esser2024scaling}, our comparative analysis reveals that the logit-normal sampler achieves superior performance compared to uniform sampling and linear scheduling alternatives.

These findings validate our parameter selections for the generative components in GCRank.

\subsection{Online A/B Test and Online Deployment of GCRank}
\label{subsec:online_deploy}
To further validate the practical effectiveness of GCRank in production environments, we deployed our framework on a large-scale takeout advertising platform and conducted a 7-day online A/B test with a 20\% traffic allocation, ensuring statistical significance through millions of daily impressions. The control group utilized our highly optimized production DLRM model that has undergone extensive refinement.

Our online deployment architecture comprises three specialized pipelines: one existing real-time serving pipeline inherited from the production DLRM system for handling online requests and streaming training, plus two newly introduced generative pipelines dedicated to updating and serving the DCE and CCE modules respectively. During serving and training, the real-time pipeline queries the DCE knowledge base and CCE embedding cache while receiving necessary updates, as detailed in Section~\ref{sec:methodology}.

Remarkably, this sophisticated architecture incurs only a minimal latency overhead, with TP99 increasing by approximately 2 ms, a modest cost that remains highly acceptable for industrial-grade advertising systems. This efficiency stems from our deliberate optimization efforts throughout the framework implementation.

GCRank achieves significant improvements across all key business metrics: CTR increased by 5.07\%, CVR improved by 4.71\%, and RPM grew by 6.42\%. All improvements are statistically significant ($p$-value $< 0.0001$), confirming both the reliability of our findings and the substantial value of GCRank in enhancing takeout contextual understanding. These results collectively demonstrate that our method delivers meaningful gains in both advertising revenue and user experience.

\begin{figure}[t]
    \centering
    \includegraphics[width=\linewidth]{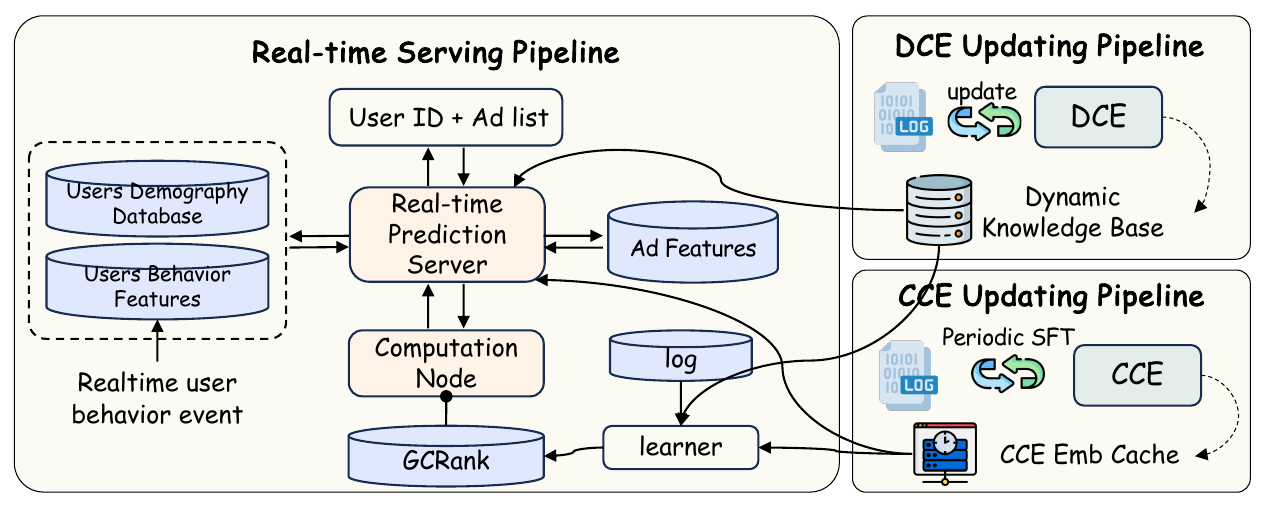}
    \caption{The online deployment architecture of GCRank, featuring three specialized pipelines for real-time serving and generative context modeling.}
    \label{fig:deployment}
\end{figure}

\section{Conclusion}
\label{sec:conclusion}

In this paper, we proposed \textbf{GCRank}, a novel generative contextual comprehension paradigm for taekout ranking model. Our key contribution lies in systematically decomposing the contextual environment into four dimensions, static, dynamic, personalized, and collective, and developing specialized modules within the Generative Takeout Context Encoder (GCE) to model each aspect. The Generative Context Fusion (GCF) module then effectively aligns and integrates these heterogeneous representations.
Extensive experiments demonstrate GCRank's state-of-the-art performance, with ablation studies confirming each component's critical contribution. Most importantly, online A/B testing on a major platform validates substantial business value through significant lifts in CTR, CVR, and RPM. This work establishes the power of generative context-aware paradigms for complex LBS-based takeout advertising scenarios.


\bibliographystyle{ACM-Reference-Format}
\bibliography{sample-base}


\end{document}